**DTIP** of MEMS & MOEMS  Stresa, Italy, 26-28 April 2006# ABOVE IC MICRO-POWER GENERATORS FOR RF-MEMS

*S. Oukassi*[1,2,3], *X. Gagnard*[1], *R. Salot*[2], *S. Bancel*[2], *J.P. Pereira-Ramos*[3]

1. STMicroelectronics, 850 rue Jean MONNET 38926 Crolles, France
2. CEA, DTNM/LCH, 17 rue des martyrs 38054 Grenoble, France
3. LECSO, UMR CNRS 758, 2 rue Henri Dunant Thiais, France## ABSTRACT

This work presents recent advances in the development and the integration of an electrochemical (chemical-electrical energy conversion) micro power generator used as a high voltage energy source for RF-MEMS powering. Autonomous MEMS require similarly miniaturized power sources. Up to day, solid state thin film batteries are realized with mechanical masks. This method doesn't allow dimensions below a few mm² active area, and besides the whole process flow is done under controlled atmosphere so as to ensure materials chemical stability (mainly lithiated materials). Within this context, Microelectronics micro-fabrication procedures (photolithography, Reactive Ion Etching…) are used to reach both miniaturisation (100x100 μm² targeted unit cell active area) and Above IC technological compatibility. All process steps developed here are realized in clean room environment.## 1. INTRODUCTION

Microelectromechanical systems (MEMS) are one of the most promising enabling technologies for developing low-cost miniaturized RF components for high-frequency applications. During the last decades, several achievements were realized in terms of technological development of RF-MEMS and integration with ICs. One important remaining integration challenge is the incorporation of the energy source on the silicon chip. Most of RF-MEMS in the literature, and mainly electrostatically driven ones require a low dc power but a high actuation voltage. The utilisation of solid state thin film batteries as a MEMS-RF power generator seems to be very adequate to the cited requirements. In fact, solid state batteries even with small footprints could provide the targeted electrical specifications. High power or voltage could then be obtained through either use of multilayered structure or a simple planar connection in parallel or in series respectively.

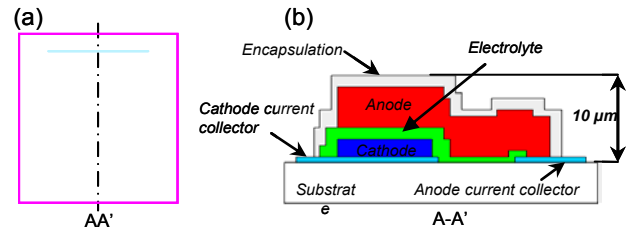

Fig. 1. Standard schematic illustration of a solid state thin film lithium battery, (a) overview and (b) cross-section

On chip integration of such devices should be compatible with standard microelectronic technology. Within this context, major efforts concern materials selection and process development. Currently, deposition processes are already compatible with standard microelectronic technology (RF magnetron sputtering, evaporation). Thin film batteries are generally realized with mechanical masking, allowing reaching dimensions as low as a few mm² of active area. On chip integration requires a miniaturization of these devices, by means of processes such as photolithography and etching. Very few studies in literature are dedicated to this subject [1-3], which is probably due to the evident difficulties associated with the processing of the materials constituting the thin film batteries.

It is the purpose of this paper to present the results of our investigations on the development of solid state thin film batteries microfabrication realized in clean room environment and using standard microelectronic technology. Results in this work concern the microfabrication process development for both cathode and solid state electrolyte thin films.

## 2. EXPERIMENTAL

### 2.1. Materials selection

Cathode and solid electrolyte materials in this study were both realized by RF magnetron sputtering deposition. Conditions are listed in Table.1.

©TIMA Editions/DTIP 2006                                                         ISBN: 2-916187-03-0

Table. 1. RF magnetron sputtering deposition conditions

| Film | Target | power density (W/cm²) | gas | Deposition pressure (mTorr) | Thickness (nm) |
|---|---|---|---|---|---|
| $V_2O_5$ | V | 4,5 | $Ar/O_2$ | 4,1 | 1000 |
| LiPON | $Li_3PO_4$ | 2,2 | $N_2$ | 15 | 1400 |

Cathode material in this study is crystalline vanadium pentoxide. $V_2O_5$ was selected for its good electrochemical performances (3V oxide, high energy storage capacity and cycle life) [4]. Another selection argument resides in the low temperature deposition process (T<200°C), which is compatible with the back-end thermal budget of current generation CMOS semiconductor technology, whereas conventional cathode materials, especially 4V oxides require higher temperature deposition and/or post annealing treatments.

A wide range of solid state materials were developed and studied as solid state electrolyte for lithium thin film batteries. For our study, glassy inorganic solid state Lithium Phosphorus Oxynitride, commonly known as LiPON [5] was selected since it presents one of the most interesting electrochemical properties (high ionic conductivity, chemical stability toward lithium up to 5.5V/Li) and also for its process compatibility (low temperature during deposition).

**2.2. Fabrication**

To provide electrical isolation for thin film batteries, $SiO_2$ and $Si_3N_4$ films were deposited respectively by thermal oxidation and low pressure chemical vapor deposition. Cathode and anode current collectors were then realized in DC sputtered Ti (50 nm)/Au (200 nm). These films were then patterned with positive resist (SJR 335, TMA 238 developer, DI rinse) and etched respectively with ($KI+I_2$) and HF solutions. Resist stripping was finally performed with HNO3 solution.

*2.2.1. $V_2O_5$ process development*

Cathode $V_2O_5$ thin film was patterned with negative photoresist (SC180 FUJIFILM electronics, Waycoat Negative Resist Developer, n-butyl acetate rinse). After development, $V_2O_5$ etching was performed in a nextral 110 RIE reactor, etch parameters were: 15 sccm $SF_6$ gas flow, 15 mTorr total pressure, and 10 W power. Spectroscopic ellipsometry characterization was used to evaluate $V_2O_5$ etch rate and conformity, and $Si_3N_4$ over-etch. Roughness Measurements (Rms and Ra) were realized with a digital instruments atomic force microscope on a 5x5 μm² window. A NEXUS FT-IR spectrometer was used to monitor residual resist evolution and $V_2O_5$ chemical stability before and after stripping step.

*2.2.2. LiPON process development*

While investigations and results on LiPON etching using ion milling were performed, we chose to present here only results obtained with wet chemical etching. Solid state electrolyte LiPON thin film was patterned, as $V_2O_5$ with negative photoresist (SC180 FUJIFILM electronics, Waycoat Negative Resist Developer, n-butyl acetate rinse). After development, LiPON etching was performed in a tetramethyl ammonium hydroxide solution (TMAH) at room temperature.

**3. RESULTS AND DISCUSSION**

**3.1. $V_2O_5$ microfabrication**

Spectroscopic ellipsometry highlights a conformal $V_2O_5$ etching over the wafer. Besides, etch rate was estimated to 28 Å/s. Figure 2 shows several $V_2O_5$ lines on $Si_3N_4$ after etching and stripping process steps. Line width is 40 μm, corresponding to mask dimensions.

Resist stripping was then performed using $O_2$ plasma RIE in the same reactor. Figure 3 shows the root mean square roughness evolution with film thickness. $V_2O_5$ presents an extremely rough surface even for reduced thickness. Next results presented here are related to 400 nm $V_2O_5$ samples showing relative less pronounced roughness. Such thickness still allows reaching electrical specifications for selected RF-MEMS.

FT-IR characterization on $V_2O_5$ after successive stripping steps shows no change of the different chemical bonds, absorption peaks of vanadyl V=O and V-O-V vibrations (respectively at 1010 and 870 cm$^{-1}$) are nearly the same for as deposited and $O_2$ plasma treated $V_2O_5$. This implies no degradation of the electrode material after the cited process step.

Residual resist on $V_2O_5$ top surface may act as lithium barrier diffusion during thin film battery charge and discharge. Thus, Residual resist detection was performed using FT-IR characterization on resist/ $V_2O_5$ samples with different durations of $O_2$ RIE (cf. Fig. 4). The identified peaks correspond well to the resist chemical composition. A total resist stripping is observed after 60 mn of $O_2$ RIE plasma and 10 mn of $O_2$ Microwave plasma. Microwave plasma is more anisotropic and seems to be more adequate since more efficient for resist stripping over $V_2O_5$ rough surface.

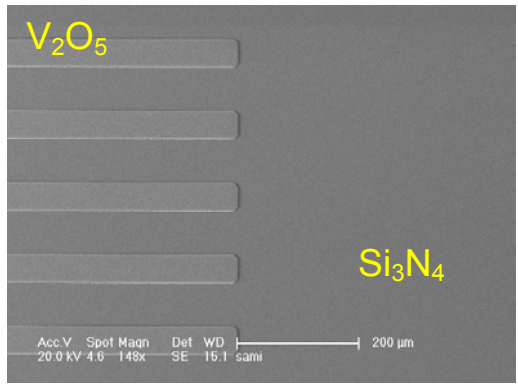

Fig. 2. SEM picture of 40 µm-width $V_2O_5$ lines after RIE etch step.

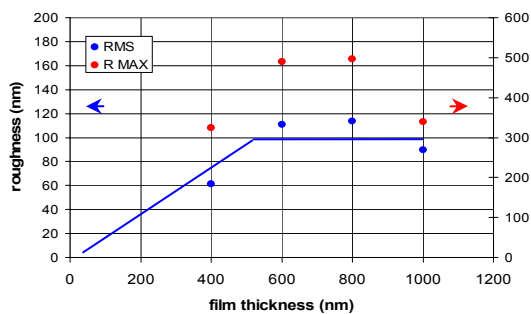

Fig. 3. $V_2O_5$ Roughness evolution with film thickness

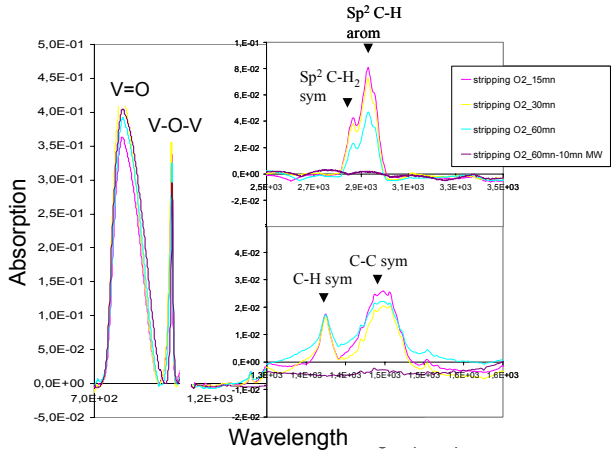

Fig. 4. FT-IR Absorption peaks evolution of $V_2O_5$ and SC 180 resist for successive plasma $O_2$ RIE: (a) Total resist stripping is observed for 60 mn $O_2$ RIE and 10 mn $O_2$ microwave plasma, (b) chemical stability of $V_2O_5$ after stripping resist step.

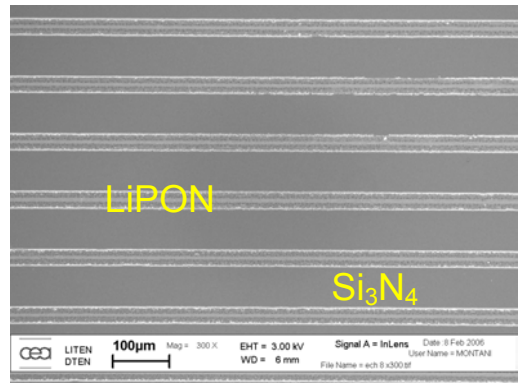

Fig. 5. SEM picture of 40 µm-width LiPON lines after wet chemical etch step.

### 3.2. LiPON Microfabrication

A total 1.4 µm LiPON thin film etch was realized in previously cited conditions over 50 s time duration.
Etch area at the edge of the lines which was estimated to 1 µm. This lateral over etch could be easily compensated during mask designing with simply adding the over etch zone to the global suited electrolyte active area. Resist stripping was then performed with $O_2$ plasma RIE.
Total resist stripping is observed after 15mn, this could be explained with the smooth LiPON surface, on the contrary of $V_2O_5$ thin film

### 3.3. Electrochemical testing

Electrochemical testing on both microfabricated $V_2O_5$ and LiPON thin film was realized separately. Solid state thin film batteries were realized both with microfabricated and mechanically masked $V_2O_5$ electrode. LiPON and lithium anode (evaporated Li) were then defined through mechanical masking, which aimed at testing only $V_2O_5$ microfabrication effects on cell electrochemical behavior. Thin film batteries were then cycled at 5µA/cm². The same approach is adopted for LiPON, which was tested in a MEM (Metal Electrolyte Metal) cell. Top and bottom electrodes were realized in RF sputtered 200 nm Pt thin films. Both microfabricated and mechanically masked LiPON cells were realized. Spectroscopic electrochemical impedance was then performed on both cell types using a Solartron impedance/gain phase analyzer, frequency ranges between $10^3$ and $10^{-2}$ Hz.

#### 3.3.1. $V_2O_5$ testing

Electrochemical cycling was performed on a (400nm microfabricated $V_2O_5$) |LiPON|Li thin film battery using the same charge and discharge characteristics as for

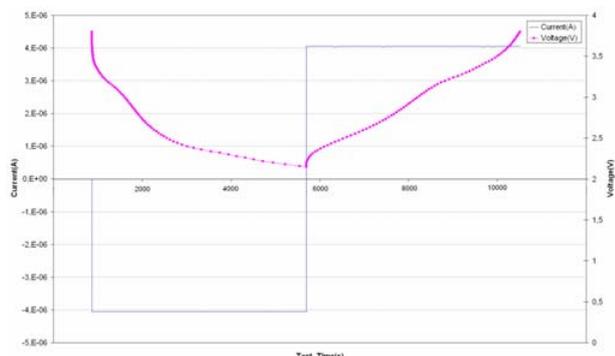

Fig. 6. Discharge-charge cycle of 400 nm microfabricated $V_2O_5$|LiPON|Li thin film battery at 5µA/cm², between 3.8 and 2.15 V.

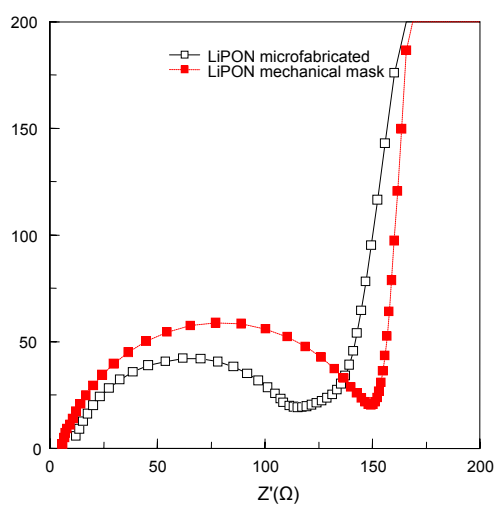

Fig. 7. Complex plane plots of Pt|LiPON|Pt cells at 297 K temperature

mechanically masked one. Electrochemical behavior is presented in figure 6. Tested cell shows a typical charge and discharge curve for the crystalline $V_2O_5$ cathode. At this stage, one can assume that the different microfabrication step does not introduce any changes on physical or chemical $V_2O_5$ properties.

*3.3.2. LiPON testing*

Figure 7 shows complex plane plots of Pt|LiPON|Pt cells performed at 297 K. The same standard plot is observed for both microfabricated and mechanically masked cells. high frequency semicircle on the plot gives an indication on the ionic conductivity σ of the LiPON electrolyte film. In fact, ionic conductivity is inversely proportional to electrolyte resistance R (estimated to semicircle diameter) following:

$$\sigma = \frac{1}{R} \cdot \frac{t_f}{S}$$

where $t_f$ is LiPON film thickness and S electrochemically cell active area. Ionic conductivities of 0.65 and 1 µScm$^{-1}$ were calculated respectively for mechanically masked and microfabricated MEM cells. The ionic conductivity of microfabricated LiPON is higher, this could be probably related to annealing steps (resist bake) during microfabrication process, and nevertheless extensive investigations about this phenomenon have not been performed.

## 4. CONCLUSION

Microfabrication process steps have been developed in clean room environment and using standard microelectronic technology so as to realize compatible Above IC thin film batteries with active areas of 100x100 µm², as a high voltage microgenerator for RF-MEMS. Electrochemical characterization results have supported the process validation of both $V_2O_5$ and LiPON active materials. Next work will refers to the anode level development taking into account the same requirements in terms of process compatibility.